\documentclass[preprint]{aastex}

\shorttitle{Gravitational lensing of S-stars} \shortauthors{Bozza
\& Mancini}

\begin{document}

\title{Observing gravitational lensing effects by Sgr\,A* with GRAVITY}

\author{V. Bozza\altaffilmark{1,2}}
 \affil{Department of Physics ``E.R. Caianiello'', University of Salerno, Italy}
 \email{valboz@physics.unisa.it}

\author{L. Mancini\altaffilmark{2}}
 \affil{Max Planck Institute for Astronomy, Heidelberg, Germany}
 \email{mancini@mpia-hd.mpg.de}

\altaffiltext{1}{Istituto Nazionale di Fisica Nucleare, Sezione di
Napoli, Italy.}%
\altaffiltext{2}{International Institute for
Advanced Scientific Studies, Vietri sul Mare (SA), Italy}

\begin{abstract}
The massive black hole at the Galactic center Sgr A* is surrounded by a cluster of stars orbiting around it. Light from these stars is bent by the gravitational field of the black hole, giving rise to several phenomena: astrometric displacement of the primary image, the creation of a secondary image that may shift the centroid of Sgr A*, magnification effects on both images. The near-to-come second generation VLTI instrument GRAVITY will perform observations in the Near Infrared of the Galactic Center at unprecedented resolution, opening the possibility of observing such effects. Here we investigate the observability limits for GRAVITY of gravitational lensing effects on the S-stars in the parameter space $[D_{\mathrm{LS}},\gamma,K]$, where
$D_{\mathrm{LS}}$ is the distance between the lens and the
source, $\gamma$ is the alignment angle of the source, and $K$ is
the source apparent magnitude in the $K$-band. The easiest effect to be observed in the next years is the astrometric displacement of primary images. In particular the shift of the star S17 from its Keplerian orbit will be detected as soon as GRAVITY becomes operative. For exceptional configurations it will be possible to detect effects related to the spin of the black hole or Post-Newtonian orders in the deflection.
\end{abstract}

\keywords{Gravitational lensing --- Black hole physics
--- Galaxy: center --- astrometry --- infreared: stars
--- instrumentation: interferometers}

\section{Introduction}
\label{Sec_1}
Understanding the nature and the physics of a black hole has
always fascinated generations of scientists as
well as ordinary people. Obviously, the first basic question is
whether there exist black holes actually. Only recently the
astronomers have been able to gain quite conclusive evidence to support a positive answer. For instance, it is widely accepted that
the central engine of an AGN is an accreting supermassive black
hole ($M_{\bullet}>10^6 \, M_{\sun}$) at the center of the galaxy,
responsible of the observed high-energy emission; stellar black
holes ($3\, M_{\sun} < M_{\bullet} < 10 \, M_{\sun}$) are
observable in close binary systems, when matter is transferred
from a companion star to the black hole; many ultra-luminous X-ray
sources are finally suspected to be intermediate-mass black holes
($M_{\bullet} > 10 \, M_{\sun}$). However, the royal evidence of
the black-hole existence resides in the innermost part of our
Galaxy, and comes from decennial near-infrared (NIR) observations
of the orbital motion of more than two dozen stars (the so-called
\emph{S-stars}) around the well-known and motionless compact radio
source, Sgr A*, which is indeed a massive black hole of about $4.3
\times 10^{6}M_{\sun}$
\citep{eckart2002,schodel2003,ghez2005a,eisenhauer2005,ghez2008,trippe2008,gillessen2009a,schodel2010}.
In particular, more than 15-years continuous monitoring of this
stellar cluster allowed to observe the whole orbit of the star S2,
which completed a full revolution around Sgr A* in 2008
\citep{gillessen2009b}. For an exhaustive review on the argument,
we refer the reader to \citet{genzel2010}.

Therefore, Sgr A* represents by far the best (and closest) place
for searching for post-Newtonian effects in strong-field regime, like for instance the relativistic motions of flaring jets and
infalling matter close to the event horizon, the deviation of
stellar orbits from Newton dynamics (relativistic periastron
shift and rotation), the precession of the angular momentum vector
around the spin of the black hole (Lens Thirring Precession), redshift variations in the spectrum of orbiting stars, etc.
General relativity clearly represents a golden key to unravel the
physical features of the black hole (spin, inclination, etc.) and
distinguish among the various (competitive) theoretical models
\citep{jaroszynski1998b,fragile2000,rubilar2001,broderick2005,weinberg2005,%
zucker2006,will2008,broderick2009,hamaus2009,merritt2009,merritt2010,angelil2011}.

Another very interesting effect expected from general relativity
is \emph{gravitational lensing}. In fact, Sgr A*, acting as a
powerful gravitational lens, is able to deflect the light rays
emitted by the S-stars from their trajectories, affecting their
measured positions and eventually generating secondary images
\citep{wardle1992,jaroszynski1998a,alexander1999,chaname2001,depaolis2003,bozza2004,nusser2004,bozza2005,bozza2009}.
However, the expected astrometric displacements of the primary
image ($\sim 20 \, \mu$as, \citealp{gillessen2009a}) and the
secondary-image luminosities ($K =20.8$ in the present known best
case, \citealp{bozza2009}) are difficult to detect and the
resolution of the most powerful modern instruments is currently
insufficient to perform such high precision astrometry and
photometry. However, the perspectives for observing S–-star
displacements and secondary images generated by the massive black
hole at the Galactic center are not so far from what the modern
technology is going to put into play. Innovative NIR
interferometry instruments are now under development at the Very
Large Telescope and Keck: they are  PRIMA \citep{delplancke2008},
GRAVITY \citep{eisenhauer2008,gillessen2010}, and ASTRA
\citep{pott2008}. These instruments have been conceived to achieve
an astrometric accuracy of $10-100 \, \mu$as in combination with
milli-arcsec angular-resolution imaging.

Here we focus on the new second-generation Very Large Telescope
Interferometer (VLTI) instrument, GRAVITY, that will be
scientifically operative at the ESO Paranal observational site
starting supposedly from 2014 \citep{eisenhauer2008}. Its
precision will allow to study the close environment of
Sgr A* in detail, testing general relativity in this so striking laboratory.
The aim of the present paper is analyze the ability of GRAVITY to
detect gravitational--lensing effects generated by the central
black hole, estimating the position and the luminosity of the
primary and secondary images of the S-stars as functions of time.

Our paper is structured as follows. In \S~2 we review the current NIR observations that have lead to the present knowledge of the Galactic center region. We also introduce GRAVITY and its expected performance. In \S~3 we examine the gravitational lensing effects by dividing the space around Sgr A* in three regions according to the GRAVITY ability to separate the images of a stellar source from the black hole. In \S~4, 5, and 6 we illustrate the respective results in the three regions. In \S~7 we draw the conclusions.

\section{NIR observations toward the Galactic center}
\label{Sec_2}
From a physical point of view, the central region of the Milky Way
is one of the most interesting zones of the Universe. Being located
at $R_{0}=8.3$ kpc from the Solar system \citep{gillessen2009a},
it is the nearest and most accessible place where to study the
interplay between a central massive black hole with the gas and the
stars in its environment. This region is also characterized by a
very high interstellar extinction (roughly 30 mag in the optical
band) along our line of sight, since it crosses the disk and
several spiral arms of the Galaxy. Nevertheless, the Galactic
center is quite transparent in the NIR bands, where the extinction
amounts to only $\approx 3$ mag. In particular, a detailed study
performed by \citet{schodel2007} showed that the interstellar
extinction in the central $\approx0.5$ pc is highly variable and
has a general minimum centered on Sgr A* of $A_{\mathrm{K}}=2.8$
\citep{eisenhauer2005}.

\subsection{Current NIR observations}
After the first observations by \emph{speckle imaging}, which have
placed the first constraints on the existence of a massive black
hole \citep{eckart1993,genzel1996,eckart1996,eckart1997,ghez1998},
the NIR observations (both imaging and spectroscopy) of the
galactic center are currently performed thanks to the adaptive
optics and laser guide technique, two very innovative technologies
introduced in the large--class telescopes ($8-10$ m) since 2002
\citep{schodel2002,ghez2003,schodel2003,eisenhauer2005,ghez2005b,ghez2008,gillessen2009a}.
Thanks to these instrumentations the imaging observations can go about 2-3 mag deeper than previous
speckle imaging observations \citep{schodel2007}.

Considering the high density of the star cluster in the
neighborhood of Sgr A*, a 8 m-class VLT unit telescope is able to
detect a faint star down to an apparent magnitude of 18.5 in the
$K$ band (around $\lambda=2.2$ $\mu$m). Instead, the limit of the
spectroscopical measurements allows to identify early- and
late-type stars as faint as $K \approx 16.5$ and $K \approx 17$,
respectively. For the brightest S-stars, it is also possible to
achieve astrometric and radial-velocity measurements with a
precision in the range $200-300 \, \mu$as and 15 km s$^{-1}$,
respectively \citep{fritz2010,gillessen2011}.

\subsection{Sgr A*}
The center of the Milky Way hosts a very exotic source, which is
variable across roughly the entire electromagnetic spectrum. It
has been detected for the first time in the radio wavelength many
years ago \citep{balick1974}, and nowadays we know that this source
is very compact and motionless, and its location coincides with
the dynamical center of the Galaxy \citep{reid2004}. Over the last
decade, Sgr A* has been fully analyzed across the electromagnetic
spectrum, at radio, sub-millimeter, NIR, and X-ray wavelengths,
revealing itself as a highly variable source especially in the
X-ray and NIR bands. Very extensive observations in the $K$ band
taken at NACO/VLT in $2004-2009$ showed that Sgr A* has a
continuously variable low-level state. The low-level emission
above 5\,mJy identifies the \emph{flaring emission state} and the low-level
emission below 5\,mJy is considered as the \emph{quiescent state},
with at least 0.5 mJy of the flux due to a faint stellar
component. The quiescent state is variable with four NIR peaks and
one X-ray peak a day. Flares that have more than 5 mJy of
variability are quite rare, roughly two per year
\citep{doddseden2011}.

\subsection{S-Stars}
The central 10 arcsec region of the Milky Way hosts a stellar
cluster, formed by old, late-type red giants, supergiants,
asymptotic giant branch stars, but also by many hot young,
early-type stars, like post-main sequence blue supergiants and
Wolf-Rayet stars. In particular, deep NIR observations of the
Galactic center (both proper-motion and Doppler measurements)
revealed $\approx 100$ stars within $r=1^{\prime\prime}$
($\approx0.04$ pc) of Sgr A*, with a remarkable concentration of B
stars, the so-called S-cluster \footnote{Unfortunately, the MPE
and the UCLA groups used different nomenclature for the stars
beloging to this cluster.}.

The existence of so many young stars in this place raises an
intriguing paradox \citep{ghez2003}. Due to the intense
gravitational field generated by the central black hole, it is
improbable that they formed in situ, but on the other hand their
age ($2-8$ Myr) is not in agreement with migration time scale.
Other alternative scenarios have been postulated, including a
mechanism of exchange capture of young binaries and the demand of
an intermediate mass black hole
\citep{gould2003,milosavljevic2004,harfst2008,lockmann2009,
madigan2009,perets2009,merritt2009,griv2010,alexander2011}.

By using the above cited high-resolution near-infrared techniques,
16 years of extensive monitoring of stellar orbits around Sgr A*,
have allowed to refine the orbital parameters of many S-stars. At the present time, the orbits of 27 S-stars have been well determined
\citep{gillessen2009a}. In a gravitational-lensing context, these
stars have already been analyzed by \citet{bozza2005,bozza2009},
who calculated all the properties of their secondary images,
including time and magnitude of their luminosity peaks and their
angular distances from the central black hole. Currently, the most
interesting case came out to be S6, since its secondary image will
reach $K=20.8$ at its peak in 2062, with an angular separation of
0.3 mas from the central black hole.

\subsection{GRAVITY} \label{Sec GRAVITY}
A big step in angular resolution is expected to come from NIR
interferometry. Actually, GRAVITY is a second-generation VLTI
instrument, specifically designed to observe highly relativistic
motions of matter close to the event horizon of Sgr A*. With a
baseline of $\approx 100$\,m, it should ensure the access to its
innermost stable circular orbit (the radius of this orbit is $30
\,\mu$as for a $4.3 \times 10^{6}M_{\sun}$ Schwarzschild black
hole), allowing astrometric detection of \emph{hotspots} orbiting
around the black hole. Thus, the high sensitivity of GRAVITY
should allow to reconstruct images with details as faint as
$K=19$, yielding a resolution of $\approx3$ mas for objects that
can be as faint as $K=18$. With this remarkable resolution,
GRAVITY should be able to catch the light of most of any very fast
orbiting stars within the central 100 mas. Its design will finish
during 2012 and it will be presumably installed in Paranal in
2014. \citep{eisenhauer2008,gillessen2010,gillessen2011}.

According to recent numerical simulations by \citet{vincent2011},
in the pure imaging mode, GRAVITY should provide an astrometric
precision of the order of, or better than, the Sgr A*
Schwarzschild radius, for a source limiting brightness of $K < 15$
and with an integration time of 100 s. However, it must be taken
into account that the astrometric precision is also very dependent
on the number of sources in the field
\citep{gillessen2010,vincent2011}.

\section{Gravitational lensing phenomena around Sgr A*}

The focus of this paper is on the effects of gravitational lensing by Sgr A* on nearby stars and how these affect their apparent properties such as position and luminosity. We will model Sgr A* as a purely Schwarzschild black hole, arguing on the validity of this hypothesis for the proposed observations all along the illustration of the results.

\subsection{Basics of lensing}
We assume a distance to Sgr A* of $D_\mathrm{OL}=8.3$ kpc and a mass of $M=4.3\times 10^6$ $M_\odot$, following \citet{gillessen2009a}. We will refer to the line connecting the observer to the lens as the optical axis. The source position is then fixed in polar coordinates around the lens by a radial distance $D_\mathrm{LS}$, a polar angle $\gamma$ taken from the optical axis, and an azimuthal angle $\phi$. The latter is irrelevant in the lensing discussion, given the assumed spherical symmetry of the lens. $\gamma$ ranges from 0 (perfect alignment of the source behind the lens) to $\pi$ (perfect anti-alignment, with the source in front of the lens). In Figure \ref{Fig 1} we illustrate the notations in the geometric set-up for gravitational lensing.

The Schwarzschild radius of Sgr A* is
\begin{equation}
R_S=\frac{2GM}{c^2}=1.27\times10^{10}\, \mathrm{m}=4.12\times10^{-7}\, \mathrm{pc}.
\end{equation}

The angular Schwarzschild radius is thus
\begin{equation}
\theta_S=R_S/D_\mathrm{OL}=10.2\, \mathrm{\mu as}. \label{thetaS}
\end{equation}

The motion of photons follows null geodesics around a Schwarzschild black hole. We can thus write an exact lens equation relating the source position $(D_\mathrm{LS},\gamma)$ to the image position $\theta$ as seen by the observer (cfr. Figure \ref{Fig 1}) \citep{FriNew,FKN,Perlick,LivRev,bozza2008,bozza2010}. To this purpose we define
\begin{equation}
A(r)=1-\frac{R_S}{r},
\end{equation}
i.e. the $g_{tt}$ component of the Schwarzschild metric or equivalently the inverse of the $g_{rr}$ component.
Considering that the impact parameter $u$ is related to the image position by
\begin{equation}
u=\frac{D_\mathrm{OL}}{\sqrt{A(D_\mathrm{OL})}}\sin\theta\simeq D_\mathrm{OL} \theta,
\label{thetau}
\end{equation}
and that the closest approach distance of the photon $r_m$ is related to this by
\begin{equation}
u^2=\frac{r_m^2}{A(r_m)}, \label{u rm}
\end{equation}
the lens equation assumes the form
\begin{equation}
2\pi\mp\gamma=\left[\int\limits_{
r_m}^{D_\mathrm{OL}}+\int\limits_{r_m}^{D_\mathrm{LS}}\right]
\frac{u}{r\sqrt{r^2-u^2A(r)}}dr, \label{Eq_1}
\end{equation}
where the upper sign holds for the primary image $\theta_+$ and the lower sign holds for the secondary image $\theta_-$ appearing on the opposite side.

The integral can be easily evaluated in terms of elliptic functions as widely reported in the literature \citep{Darwin}.

The magnification of a gravitational lensing image can be calculated by taking the ratio of an angular displacement in the observer sky and the corresponding displacement in the source surface orthogonal to the emission direction of the photons. Following \citet{bozza2010}, and expressing the result in a synthetic form, we have
\begin{equation}
\mu=\frac{D_\mathrm{OS}^2}{D_\mathrm{LS}^2 D_\mathrm{OL}^2}\frac{u}{\sin\gamma} \left( \sqrt{1-u^2/D_\mathrm{LS}^2}\frac{du}{d\gamma} + u \frac{du}{dD_\mathrm{LS}} \right) \label{mu}
\end{equation}
where $D_\mathrm{OS}=\sqrt{D_\mathrm{OL}^2+D_\mathrm{LS}^2 +2D_\mathrm{OL}D_\mathrm{LS}\cos \gamma}$ is the distance from the observer to the source. The derivatives $du/d\gamma$ and $du/dD_\mathrm{LS}$ can be evaluated numerically using the lens equation. Note that for small impact parameters compared to the source distance ($u\ll D_\mathrm{LS}$) we recover the magnification formula by \citet{ohanian1987}. In the context of black hole lensing, this is sometimes referred to as thin lens approximation.

Gravitational lensing effects are maximal for $\gamma\simeq0$ (standard lensing), with a second maximum for the secondary image occurring for $\gamma=\pi$ (retro-lensing). This is reflected by the $\sin\gamma$ in the denominator of the magnification formula.

In the weak field limit, the integral of Eq. (\ref{Eq_1}) simplifies and the lens equation becomes
\begin{equation}
\pm\gamma=\frac{r_m}{D_\mathrm{OL}}+\mathrm{arcsin } \frac{r_m}{D_\mathrm{LS}}-\frac{R_S}{r_m}- \frac{R_S(2D_\mathrm{LS}+r_m) \sqrt{\frac{D_\mathrm{LS}^2}{r_m^2}-1}}{D_\mathrm{LS}(D_\mathrm{LS}+r_m)} \label{WF LEQ},
\end{equation}
where we have just saved the lowest order term in $r_m/D_\mathrm{OL}$, whereas we allow $D_\mathrm{LS}$ to become comparable to $r_m$.

The relation between the minimum distance $r_m$ and the impact parameter $u$ can be written to first order in $R_S$ as
\begin{equation}
r_m=u-\frac{R_S}{2}+o(R_S).
\end{equation}

In addition to the weak field approximation, the thin lens approximation requires $D_\mathrm{LS}\gg r_m, u$. This automatically implies small angles, so that we can set $\gamma=D_\mathrm{OS}\beta/D_\mathrm{LS}$, where $\beta$ is the observed angular position of the source if there were no lens (see Figure \ref{Fig 1}). The small angles weak field lens equation is then obtained
\begin{equation}
\beta=\theta-\frac{\theta_E^2}{\theta}, \label{WFSA LEQ}
\end{equation}
where
\begin{equation}
\theta_E=\sqrt{\frac{2R_S D_\mathrm{LS}}{D_\mathrm{OL}D_\mathrm{OS}}}
\end{equation}
is the so-called Einstein angle.

Summing up, the full general relativistic treatment without approximation is given by Eq. (\ref{Eq_1}). The weak field approximation is given by Eq. (\ref{WF LEQ}) and the weak field + thin lens approximation is given by Eq. (\ref{WFSA LEQ}). Note that if we want to test the sensitivity of gravitational lensing to Post-Newtonian orders of the black hole metric, we must compare the results of Eq. (\ref{WF LEQ}) to Eq. (\ref{Eq_1}). In fact, in many cases the differences between Eq. (\ref{Eq_1}) and Eq. (\ref{WFSA LEQ}) are not imputable to strong field effects but just to the thin lens/small angles approximation.

Nevertheless, it is useful to recall the simple analytical expressions of images and magnification obtained by Eq. (\ref{WFSA LEQ}), since they provide a useful guide to the expected gravitational lensing effects. We have
\begin{eqnarray}
&& \theta_\pm=\frac{1}{2} \left(\beta\pm\sqrt{\beta^2+4\theta_E^2}\right) \label{thetaw}\\
&& \mu_\pm=\frac{1}{2} \left(\frac{\beta^2+2\theta_E^2}{\beta\sqrt{\beta^2+4\theta_E^2}} \pm 1\right) \label{muw}
\end{eqnarray}

The primary image is always outside the Einstein ring with radius $\theta_E$, while the secondary image is always inside the ring. Both images approach the Einstein ring as $\gamma$ approaches 0, merging into a single degenerate ring image with radius $\theta_E$ for $\gamma=0$.

\subsection{Strong field and spin corrections} \label{Sec Strong}

Sgr A* offers a unique chance to test GR in a strong field regime.
It is thus of primary importance to understand to what extent a weak field Schwarzschild treatment based on Eq. (\ref{WFSA LEQ}) is sufficient to explain all the phenomenology within reach of GRAVITY and at which point the full GR treatment of Eq. (\ref{Eq_1}) is really necessary. This will also indicate where to look for new tests of GR based on lensing observations.

\subsubsection{Black holes in alternative gravity theories}

Most alternative theories of gravity only become effective in the neighborhood of the event horizon, by predicting space-time metrics deviating from Schwarzschild in such an extreme regime. These deviations are presumed to assume sizeable proportions for Sgr A*, which is the closest massive black hole to us. At the same time, gravitational lensing provides one of the simplest ways to test these alternative theories once the metric is known. For this reason, gravitational lensing by Sgr A* is invoked as the natural astrophysical test in these theoretical studies (see e.g. \citet{bozza2010,LivRev} and references therein). This class of alternative metrics includes black holes with scalar fields, string theory black holes, braneworld black holes, wormholes, Born-Infeld gravity black hole.

All these metrics predict appreciable effects for the higher order images, i.e. those images generated by photons looping around the black hole once or more times. However, these higher order images require resolutions of the order of $\mu$as or better and exceptional alignments to be detected, since they are extremely demagnified. In the context of lensing of S-stars, higher order images were examined by \citet{bozza2004,bozza2005} for the Schwarzschild metric, where it t is shown that they are always fainter than $K=30$ for all known S-stars. A first direct attempt to find effects of alternative gravity theories in lensing of S-stars has been lead by \citet{Binnun2010,Binnun2011}, who considered a tidal Reissner-Nordstrom metric coming from Randall-Sundrum II theories. The perturbations are larger for images forming very close to the black hole, but become very small for secondary images forming far enough from the black hole  ($\Delta K\sim 0.01$ for S6).

Summing up, the common feature of all these metrics is that they reproduce the Newtonian limit far from the black hole, while they differ from Schwarzschild in their Post-Newtonian expansions. The position and the magnification of the images can be calculated analytically beyond the weak field limit of Eqs. (\ref{thetaw})-(\ref{muw}) for a generic Post-Newtonian metric, by an expansion in powers of the Schwarzschild radius \citep{Ebina,Lewis,Sereno,KeePet}. The relative correction is of order
\begin{equation}
\frac{\delta \theta_\pm}{\theta_{\pm}}\approx \frac{\delta \mu}{\mu}=\frac{\theta_S}{\theta_\pm}.
\end{equation}

It therefore takes effect only for light rays passing very close to the event horizon. This happens for sources very close to the black hole (e.g. on the accretion disk) or for secondary images generated by sources far from the optical axis. In the latter case, the secondary images also become very faint and difficult to distinguish from the lens. For this reason, S-stars do not seem to be quite ideal candidates to study strong field effects. However, we will see that there are regimes in which higher order deflection terms may become marginally accessible. Therefore, for each observable, we will indicate the region in which a source should be in order to yield sizeable Post-Newtonian effects.

\subsubsection{Alternatives to the black hole hypothesis}

Alternatives to the standard black hole paradigm come from boson \citep{TorCapLam}, fermion stars \citep{Viollier} and dark matter concentrations \citep{TsiVio}. While these objects are built so as to reproduce the observed spectrum of the black hole accretion disk \citep{guzman}, their extended nature leaves a fundamental signature in gravitational lensing, since the deflection angle no longer grows when the impact parameter is decreased below the physical radius of these stars. However, the restrictions imposed by their construction scheme makes this difference arise only for light rays approaching closer than a few Schwarzschild radii. So, once more a source very close to the central object is needed.

\subsubsection{Modified gravity}

Among the many attempts to solve the dark matter/dark energy problems by modifying GR, the MOND theory is the most famous \citep{milgrom}, thanks to its recent relativistic version by \citet{bekenstein}, called TeVeS. In this case, gravity is modified at small fields, when the gravitational acceleration drops below a new fundamental constant $a_0\sim 10^{-10}$ m/s$^2$. Gravitational lensing by point-like objects in this context was studied by \citet{chiu}. It turns out that deviations from Schwarzschild lensing arise for impact parameters comparable or larger than the scale
\begin{equation}
r_0=\sqrt{G M/a_0} \simeq 77\; \mathrm{pc}.
\end{equation}

These effects would thus be important for far background bulge stars lensed by Sgr A*, in which case the competing effect of the stellar mass of the inner clusters that would sum up with the black hole lensing would probably make a clear detection of TeVeS lensing very difficult.

In the case of S-stars, whose distance from Sgr A* is below the pc, TeVes corrections are largely negligible (order nas at most in the astrometric shifts).

\subsubsection{Black hole rotation}

Rotating black holes are described by the Kerr metric, characterized by the spin parameter $a$, corresponding to the black hole specific angular momentum, ranging from 0 (Schwarzschild) to $M$ (extremal rotation). In Kerr black holes, the point-like caustic of the single lens becomes a small astroidal caustic, shifted from the optical axis by $a R_S/2M$ in the west direction (if we identify north as the direction of the black hole spin projected on the sky) \citep{SerDeL}. This means, that in a first approximation the Kerr black hole works as a Schwarzschild black hole slightly shifted from its position \citep{AsaKas,AsaKasYam}. If the center of Sgr A* is pinpointed to very high accuracy (below the angular extension of the Schwarzschild radius), then all gravitational lensing effects will work as if the black hole were displaced by $aR_S/2M$. Since many quantities diverge as the inverse of the distance of the source to the optical axis ($D_\mathrm{LS}\gamma$ in our notations), even such a small displacement of the caustic may induce dramatic effects. For this reason, we will also indicate the limits of sensitivity to the spin for each observable we are going to examine.

\subsubsection{Thin lens approximation break-out}

The last possible break-out of the weak deflection lens equation (\ref{WFSA LEQ}) arises when the source is very badly aligned ($\gamma \gg 1$). Then the primary image cannot be calculated using the so-called thin-lens approximation, which in this context consists in sending the integration limits to infinity in Eq. (\ref{Eq_1}). In this case, in fact, $D_\mathrm{LS}$ becomes comparable to the impact parameter $u$.

\subsubsection{Image distortion}

Finally, we also note that it would be extremely improbable to see any kind of image distortion of the stars around Sgr A*. In fact this is possible only for alignment angles $\gamma$ of the order of the source angular radius, which in the case of S-stars is of the order of ns of arc. For comparison, in Figure \ref{Fig 2} we report the positions of S27 and S6 at the best alignment epoch of their orbit. These two stars, which represent the best known cases, at most get at $\gamma\simeq 3^\circ$. Therefore, we will always consider the images of the S-stars as effectively point-like for GRAVITY and thus described by a PSF. For the same reason, the finite size of the Kerr caustic in a rotating black hole does not lead to any observable effects.

\subsection{Three regimes where to look for lensing in the Galactic center}
Gravitational lensing generates two images of a given source star. These images might be confused or not within the PSF of Sgr A* itself, depending on the angular position of the images compared to GRAVITY resolution limit of 3 mas. We can therefore distinguish three different regimes for gravitational lensing, occurring in three different regions of the plane $(\gamma,D_\mathrm{LS})$.
\begin{itemize}
\item{Region I: No images are resolved from Sgr A*. A single blend appears containing both images of the source star along with the radiation from the central black hole.}
\item{Region II: The primary image is resolved separately while the secondary image is still confused with Sgr A*.}
\item{Region III: Both images are distinguished from Sgr A* and can be studied separately.}
\end{itemize}

Note that $|\theta_-|<|\theta_+|$ so that it is not possible to have the secondary image resolved with the primary unresolved.

The boundaries of the three regions can be easily calculated using the lens equation (\ref{Eq_1}) by imposing $\theta_\pm=\theta_{thr} \equiv 3$ mas in the case of GRAVITY and solving for $\gamma$ as a function of $D_\mathrm{LS}$. In Figure \ref{Fig 2} we can see the shapes of the three regions in the plane $(\gamma,D_\mathrm{LS})$.

Region I covers the immediate neighborhood of Sgr A*, when the source is too close to the black hole to generate a distinguishable primary image. Using the weak field/thin lens formalism (Eq. \ref{WFSA LEQ}), an approximate expression for its boundary is
\begin{equation}
\gamma_{I-II}=\frac{D_\mathrm{OL}+D_\mathrm{LS}}{D_\mathrm{LS}}\theta_{thr}-\frac{2R_S}{D_\mathrm{OL}\theta_{thr}}. \label{Boundary I-II}
\end{equation}

The tip of region I at $\gamma=0$ extends up to
\begin{equation}
D_\mathrm{cr}=\left(\frac{2R_S}{D_\mathrm{OL}^2\theta_{thr}^2}-\frac{1}{D_\mathrm{OL}}\right)^{-1}=0.018\, \mathrm{pc}.
\end{equation}

Beyond this point the primary image will always be resolved by GRAVITY whatever the alignment angle $\gamma$. The weak-deflection analytical expression tracks the exact boundary extremely accurately, as $\theta_\mathrm{thr}\simeq 300 \theta_S$.

The possibility to observe both images as separate objects from Sgr A* is restricted to sources within a cone starting at the tip of region I. The analytical expression coming from the weak deflection approximation is just the opposite of Eq. (\ref{Boundary I-II})
\begin{equation}
\gamma_{II-III}=-\frac{D_\mathrm{OL}+D_\mathrm{LS}}{D_\mathrm{LS}}u_{thr}+\frac{2R_S}{D_\mathrm{OL}u_{thr}}. \label{Boundary II-III}
\end{equation}
For sources at large distances we have $\gamma_{II-III}\rightarrow 0.39^\circ$.

In the next three sections we will closely examine the three regions just defined, pointing to possible interesting observational perspectives with GRAVITY, considering either photometric or astrometric detection channels.

\section{Region I}
In region I of the parameter space (Figure \ref{Fig 2}),
the primary image of a generic stellar source is within 3 mas
from the black hole. Then, it cannot be resolved from the radiation emitted by Sgr A* and forms a single blend along with the secondary image. Obviously, in this situation it is difficult to extract pure gravitational lensing signals on the lensed star, as everything is mixed with the variable flux from Sgr A*.

\subsection{Astrometry}

Let us consider astrometric variations first. The centroid of the blend composed of Sgr A*, primary image and secondary image of the lensed star can be calculated by evaluating the position of the images by Eq. (\ref{Eq_1}) and their magnification from Eq. (\ref{mu}). Of course, the weight of the two images with respect to Sgr A* depends on the relative fluxes in the $K$-band. By considering that a magnitude of $K=0$ corresponds to 667 Jy and that the extinction is $A_{\mathrm{K}}=2.8$ \citep{eisenhauer2005}, Sgr A* has an apparent magnitude of $K\approx 17$ in its quiescent state, corresponding to $1.1$ mJy. As mentioned before, this flux is highly variable even within the same day. However, for the sake of simplicity, we have taken $K_\mathrm{Sgr A*}=17$ for the calculation of the centroid shift. As for the stellar source, typical values of S-stars range from $K=13$ to $K=17$. This also means that if gravitational lensing effects take place, the centroid will be typically dominated by the primary image shift, as Sgr A* is relatively faint.

In Figure \ref{Fig 3} we show the expected astrometric shift of the centroid in region I due to gravitational lensing effects for a source with $K=15$. In practice, we are comparing the centroid position of the blend including Sgr A* and the two lensed images, with the centroid position obtained with Sgr A* and the unlensed source. As expected, the shift is higher for $\gamma\rightarrow 0$, i.e. for better alignments. Considering that the astrometric sensitivity of GRAVITY is of the order of $30$ $\mu$as, gravitational lensing effects would be detectable for a large portion of this region up to $\gamma=30^\circ$. This anticipates the results of the astrometric effects in region II, to be discussed in the next section, where the primary image is resolved and things are much cleaner. In region I, instead, the variability of Sgr A* would add some noise on the astrometric measurements. A reliable extraction of the gravitational lensing astrometric shift would require follow-up observations of the whole transit of the source behind Sgr A*, so that the noise due to Sgr A* can be efficiently subtracted.

Strong field and spin effects on the total centroid shift in region I are tiny, staying below the $\mu$as even for sources closer than $10^{-4}$ pc.

\subsection{Photometry}

The second channel for the detection of gravitational lensing effects comes from photometry. In this case as well, the variability of Sgr A* adds noise to the flux variations of the primary and secondary images. We have thus fixed a threshold for photometric detection of lensing magnification effects of $\Delta K=0.1$. The time-scale of lensing variations depends on the radius and the eccentricity of the orbit of the source S-star. For example, the transit of a star on a circular orbit at a distance $D_\mathrm{LS}=0.01$ pc behind the black hole in the region within 3 mas of Sgr A* lasts about 100 days, scaling with $\sqrt{D_\mathrm{LS}}$. As this time is much larger than the scale of the variability of Sgr A*, the two phenomena can be disentangled with suitable follow-up observations.

Figure \ref{Fig 4} shows region I split in two zones: the left one where the photometric variation due to gravitational lensing is detectable and the right one where the magnification effect is too small. The three curves track this boundary for three values of the source magnitude in $K$-band (13, 15, or 17). Note that an S-star with $\gamma \gtrsim 10^{\circ}$ will not induce any perceptible flux variation even if it were more brighter than S2. Based on our knowledge of the orbits of several S-stars, we can say that GRAVITY will always be able to separate the primary image from Sgr A*. This means that at present we do not know S-stars falling in region I. However, GRAVITY will likely discover more stars that are too close to be detected with present facilities. It appears a reasonable guess that some of them will enter the regime described in this section along their orbit. If this is the case, their gravitational lensing effects can be studied along the lines traced in this section.

Interestingly, the photometric magnification of a stellar source provides a first channel for the detection of the spin of the black hole. In Fig. \ref{Fig 4} the dotted line bounds the region in which the displacement of the Kerr caustic discussed in Section \ref{Sec Strong} modifies the magnification in such a way that the total observable flux changes by more than $\Delta K=0.1$. Of course, in order to distinguish the spin signal, we need to know the relative positions of Sgr A* and the source S-star to very high accuracy from previous observations. Contrarily to the spin, Post-Newtonian terms in spherically symmetric black holes do not give rise to detectable effects.

We finally note that the threshold precision of $10\%$ that we have requested here has been chosen by considering the scatter in the data collected with the Naos-Conica (NACO) system mounted at the fourth unit telescope Yepun of the VLT (see over Figure 8 of
\citealp{gillessen2009a})\footnote{The question of how accurate the
$K$ values are, is actually difficult, because is not easy to
understand how the systematic uncertainties affect the
measurements. For instance, the extinction around Sgr A* is not
constant spatially but at the contrary, as showed by the accurate
extinction map of \citet{schodel2007}, is very patchy. This means
that stars that hang around the Galactic center experience higher
and lower extinctions, which modulates the measured fluxes. One of
the possible explanation of the S2 flux increasing observed in
2002, when the star passed its pericenter, is just that (Gillessen
2009, private communication).}. We are quite confident that GRAVITY should be able to
achieve a similar precision in measurements of the same type.

\section{Region II} \label{Sec Reg II}
In this region, the primary image of the source star is detached
from the black hole, whereas the secondary image is not. Obviously, this regime includes the no-lensing limit with $\beta \gg \theta_E$, in which the primary image is basically unlensed and the secondary image becomes negligible. All known S-stars fall in this region, so that GRAVITY will be able to resolve their primary image all along their orbits (in current VLT observations, instead, S2 and S6 used to be unresolved from Sgr A* for some periods).

In this regime, we can look for four different gravitational lensing effects: astrometric shift of the primary image, magnification of the primary image, centroid shift of the blend composed of Sgr A* and the secondary image, magnification effects on the same blend.

\subsection{Astrometry on the primary image}

The astrometric shift on the primary image due to gravitational lensing by the central black hole has already been studied in detail by \citet{nusser2004}. These authors could not envisage the breakthrough in astrometric accuracy promised by GRAVITY, so that their conclusions were not very optimistic for the observational perspectives of such an effect in a short term.

For this reason, it is interesting to revisit the expected astrometric shift in the source coordinate space in the light of GRAVITY expectations. Figure \ref{Fig 5} shows contour lines for the astrometric shift at several values. Strikingly, many of the known S-stars would have experienced a measurable astrometric shift during their passage behind Sgr A*. Even at alignment angles as large as $\gamma=35^\circ$ the astrometric shift is of the order of 30 $\mu$as. This encourages to look for astrometric shift effects on known S-stars as soon as GRAVITY becomes fully operational. To this purpose, in Figure \ref{Fig 6} we show the position of the S-stars in the $(\gamma,D_\mathrm{LS})$ plane in the years from 2014 to 2024. We see that S17 will have an appreciable shift for a few years, while several S-stars may have a marginally detectable shift. Interesting targets will be S29 and S2, which will make rapid incursions in this plot as they will pass through their periapse.

Figs. \ref{Fig 5} and \ref{Fig 6} clearly show that the shift is practically independent of the source distance. This statement can actually be supported by a simple analytical derivation. In fact, since light rays of the primary image pass very far from the event horizon, the weak field lens equation (\ref{WF LEQ}) gives the same results as the exact lens equation (\ref{Eq_1}) to very high accuracy (the difference at most reaches 5 $\mu$as for $\gamma \rightarrow 0$). Conversely, at large alignment angles $\gamma$, the small angle approximation (\ref{WFSA LEQ}) fails. Starting from Eq. (\ref{WF LEQ}), solving perturbatively for $u$ in powers of $R_S$ and taking the limit $D_\mathrm{OL}\gg D_\mathrm{LS}$, we find
\begin{equation}
\Delta \theta=\frac{R_S\cos^3 \gamma/2}{D_\mathrm{OL}\sin\gamma/2}, \label{astro2}
\end{equation}
which is independent of $D_\mathrm{LS}$.

Proceeding in a similar way from the small angles thin lens equation (\ref{WFSA LEQ}), we get
\begin{equation}
\Delta \theta\simeq \frac{2R_S}{D_\mathrm{OL}\gamma},
\end{equation}
which coincides with the small $\gamma$ expansion of Eq. (\ref{astro2}), but gives an error of order $10\%$ at $\gamma=30^\circ$ already and thus cannot be used for accurate shift calculations on S-stars.

We finally note that this gravitational lensing shift should be taken into account in any accurate orbit determination undertaken from GRAVITY data. This effect will help setting further constraints on the black hole mass, the mass distribution around it and the orbital parameters of S-stars. As anticipated, Post-Newtonian and spin effects for the primary image shift in region II are negligible.

\subsection{Astrometry on the secondary image}

In region II the secondary image is still blended with the emission from Sgr A*. As the source aligns with the black hole, the secondary image gets brighter and moves away from Sgr A*. Both these effects contribute to shift the centroid of the blend in the direction opposite to the primary image. Obviously, the centroid shift is determined by the balance between the luminosity of Sgr A* and that of the source S-star.

Similarly to what we did for region I, we calculate the centroid shift of this blend. In Figure \ref{Fig 7} we show the detection limits of GRAVITY (set to 25 $\mu$as) for three different values of the source magnitude in the $K$-band. Interestingly, S27 and S6 at their closest alignment would have fallen inside the detection zone if they were more luminous (S6 has $K=15.4$ and S27 has $K=15.6$). No other known S-stars generate secondary images prominent enough to be detected by GRAVITY.

Post-Newtonian effects on the secondary image are larger since this image is formed by light rays passing close to the event horizon. However, in the region of interest in Figure \ref{Fig 7}, the difference between the exact centroid shift and the shift calculated by the weak field formalism stays below 1 $\mu$as. The spin effects reach at most 2 $\mu$as, one order of magnitude below GRAVITY's sensitivity. In principle, using a very large sample of observations of different stars whose orbit is very well known, it might be possible to estimate the spin and Post-Newtonian terms by a careful statistical analysis. So, this is an interesting possibility opened by a measure that is in principle feasible with GRAVITY, provided that some S-stars pass through this region. Obviously, the lensing signal is confused here with the variability of Sgr A*, which needs to be subtracted with long-period observations.

\subsection{Photometry on the primary image}

The magnification of the primary image is another possible channel for detecting gravitational lensing. It is inversely proportional to the alignment angle (as long as the source radius is negligible), so that it requires quite accurate alignments. The expected magnification for the known S-stars is very low (we estimated that the best case is
$\mu=1.01$ for S6) and so surely undetectable also by GRAVITY. If we consider a detection threshold of $\Delta K=0.1$, the candidate sources must be on the left of the dashed line in Figure \ref{Fig 8}, i.e. within $\gamma=1^\circ$. It is a small window for this channel, but not impossible to achieve for S-stars to be discovered yet. By crossing astrometric and magnification measurements it is possible to get even more stringent constraints on the lens nature.

\subsection{Photometry on the secondary image}

The photometry on the secondary image is complicated by the blending with Sgr A*. The threshold at $\Delta K=0.1$ allows for a relatively wide detection region depending on the source magnitude. In Figure \ref{Fig 8} we see its boundary for three values of the magnitude of the S-star. With respect to astrometry, the region extends at larger angles at small $D_\mathrm{LS}$, while it becomes thinner at larger $D_\mathrm{LS}$. The difference between the exact and weak field values is smaller than in astrometry and is thus not shown in the figure. Also in this case, no known S-stars fall within the detection limits of GRAVITY.

Note that the detection limit for $K=17$ practically coincides with the limit for the primary image. This is a trivial consequence of the fact that the magnitude of Sgr A* in its quiescent state has been taken to be $K=17$, i.e. equal to the unlensed source magnitude.

Post-Newtonian and spin contributions to photometric amplification remain small both for the primary and for the secondary image, starting to be significant only for $\gamma<1'$.

\section{Region III}

In this region, the two images of the source star appear distinct from Sgr A*. Obviously, this is the most favorable case for a gravitational lensing detailed study. The mass distribution and the distances can be constrained by simultaneous measurements on the primary and the secondary image position and magnification, so that the greatest benefits for a study of the central environment is achieved. Unfortunately, region III is a cone with aperture angle of $\gamma=0.39^\circ$ only, so that the chances for such a good alignment are quite low.

Furthermore, we still have to discuss the actual detectability of the secondary image, which is always much fainter than the primary. Following our discussion of GRAVITY performance in Section \ref{Sec GRAVITY}, we consider a threshold magnitude of $K=19$ for the secondary to be clearly detected. Depending on the source intrinsic luminosity, in Figure \ref{Fig 9} we draw three detection limits corresponding to this threshold. These limits demand an even higher degree of alignment for the secondary image to be effectively visible.

In the weak field/thin lens formalism, these curves can also be calculated analytically by imposing that the magnification of the secondary image (\ref{muw}) is above the threshold $\mu=10^{-0.4(19-K)}$, where $K$ is the magnitude of the source. We obtain
\begin{equation}
\gamma=\sqrt{\frac{2R_S}{D_\mathrm{LS}}\frac{(1+2\mu)-2\sqrt{\mu(1+\mu)}}{\sqrt{\mu(1+\mu)}}}.
\end{equation}

Consider that at a given distance $D_\mathrm{LS}$, the radius of the detection zone scales as $D_\mathrm{LS} \gamma\simeq \sqrt{D_\mathrm{LS}}$, i.e. the detection zone effectively grows with the distance, contrarily to the impression given by Figure \ref{Fig 9}.

Another interesting aspect is that throughout region III the contribution of Post-Newtonian and spin effects to the astrometric shift is relatively high: 5 $\mu$as for the primary image and up to 10 $\mu$as for the secondary image. These values are still below GRAVITY sensitivity. However, as already noted before, a large number of observations can be used to study these effects with a statistical approach.

We finally remind the estimates of previous works on the number of bulge stars that should be lensed by Sgr A* at a given time with a threshold of $K=23$ for the secondary image is around 10 \citep{jaroszynski1998a,alexander1999,chaname2001}. Here we consider a more conservative threshold for GRAVITY, since we are considering its specific use for interferometry.

\section{Discussion and conclusions}
Thanks to the coherent combination of the four 8-m Unit Telescopes
and the four 1.8-m Auxiliary Telescopes, which can be relocated on
30 different stations, the VLTI is the only interferometer that
allows to perform very high-precision photometry, with mas angular
resolution, and a total collecting area of 200 m$^{2}$. In
particular, the size of the total field of view combining the four
8-m Units is of $2^{\prime\prime}$. Since no other interferometer
of such class are currently in the design phase, the VLTI
will be at the leading-edge still for many years to come. After
the VINCI, MIDI \citep{leinert2003}, AMBER \citep{petrov2007} and
the visitor instrument PIONIER \citep{lebouquin2011}, GRAVITY
arises as the second generation VLTI instrument and will perform
high precision narrow angle astrometry as well as interferometric
imaging in the $K$-band (2.2 $\mu$m).

Taking into account the capabilities of GRAVITY, we have analyzed the possibility to observe gravitational lensing effected generated by the massive black hole, located in the geometrical center of the Milky Way, on the S-stars orbiting around it.

We have divided the parameter space into three regions (Figure \ref{Fig 2}), depending on the resolution of the primary and the secondary image from the black hole and we have carefully investigated astrometric and photometric channels for the detection of gravitational lensing effects.

\begin{figure}
\centering{\includegraphics{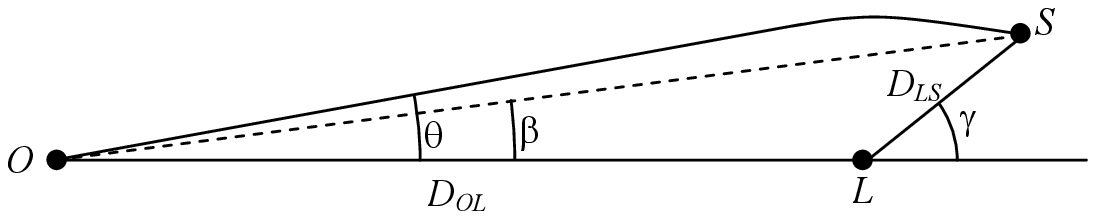}}
 \caption{Generic lensing configuration.}
 \label{Fig 1}
\end{figure}

\begin{figure}
\centering{\includegraphics{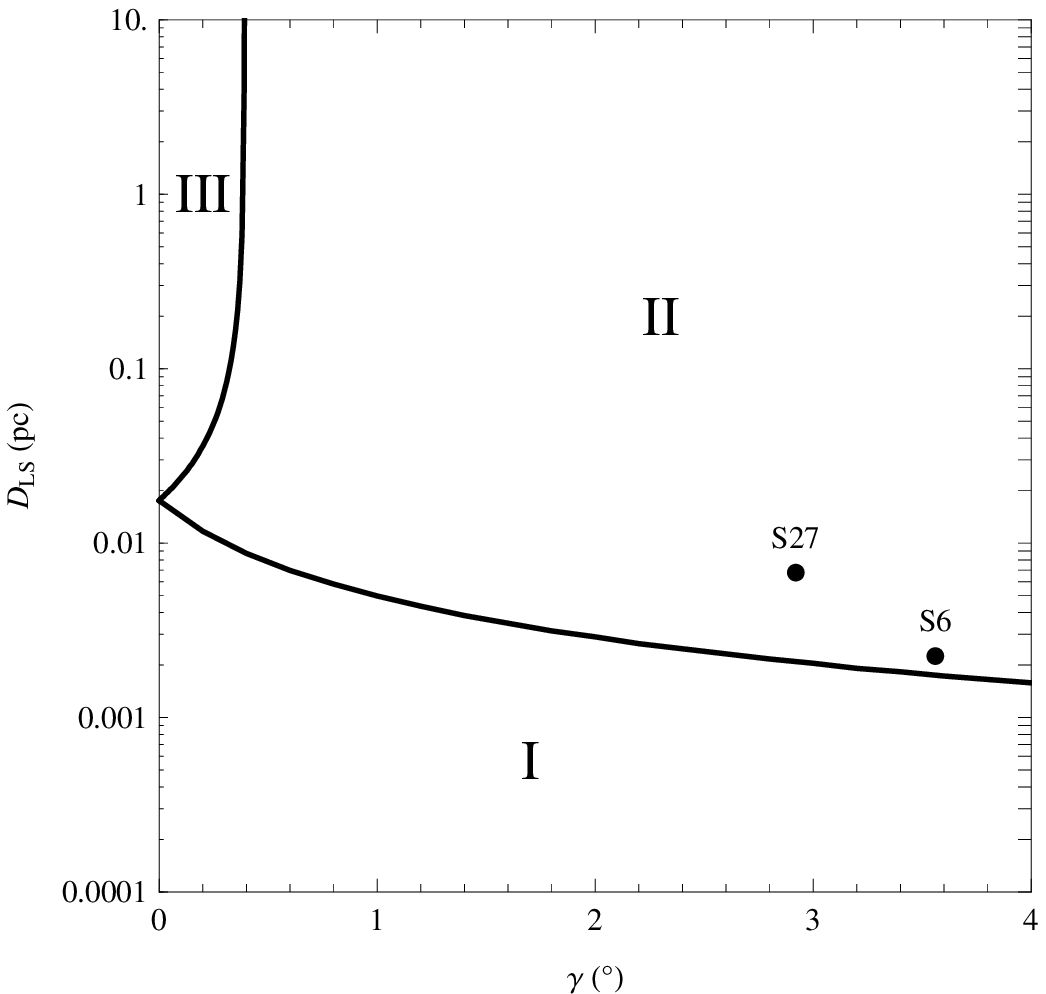}}
 \caption{The source-star parameter space examined.
 We can identify three regions where different gravitational lensing regimes take place according to the expected resolution of GRAVITY (3 mas). In region I both images are unresolved from Sgr A*. In region II the primary image is separated from Sgr A* while the secondary is not. In region III both images are resolved. The positions of the stars S6 and S27, at their best
 alignment along the optical axis, are reported too.}
 \label{Fig 2}
\end{figure}

\begin{figure}
\centering{\includegraphics{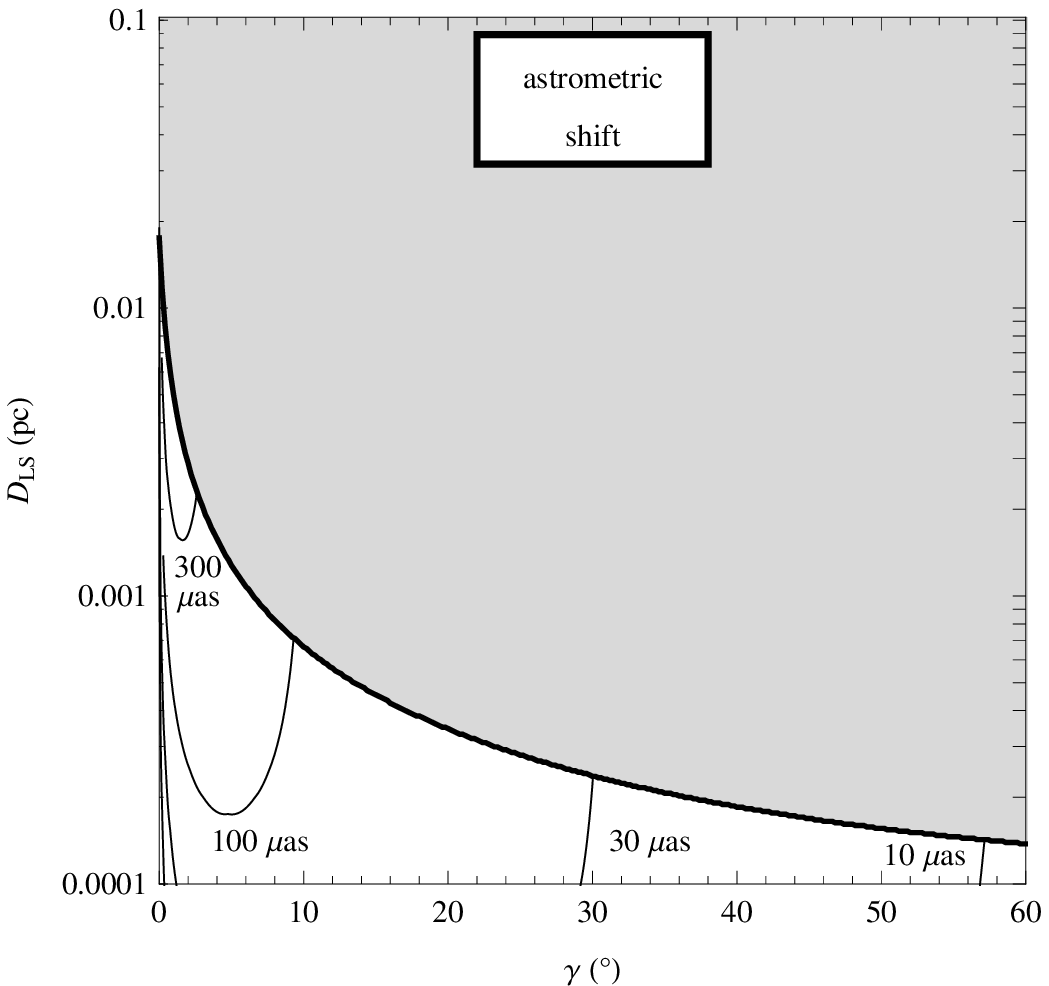}}
 \caption{Region I: astrometric shift of the centroid of the blend composed of Sgr A* and the two lensed images of a source with $K=15$. The shift is calculated with respect to the position of the centroid of the blend composed of Sgr A* and the unlensed source. For Sgr A* we are assuming $K=17$.}
 \label{Fig 3}
\end{figure}

\begin{figure}
\centering{\includegraphics{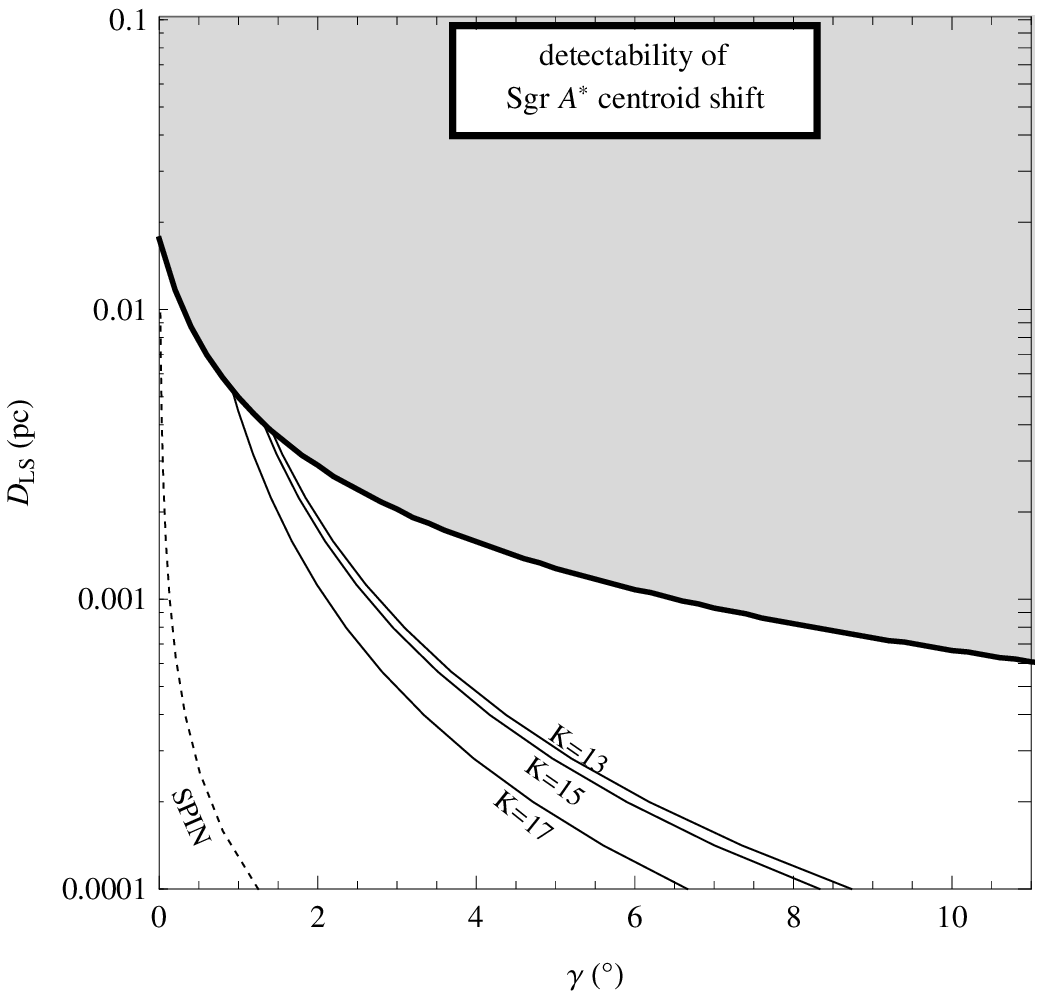}}
 \caption{Details of region I, where both the primary and the secondary image of a generic S-star
 are confused with Sgr A*.  The three curves (for a source star of $K=13,15,17$ respectively) bound the inner zone in which gravitational lensing magnification effects lead to an excess of $\Delta K=0.1$ in the blend composed by Sgr A* and the two images of the lensed star. The dotted line bounds the inner region in which the measure of the total flux is sensitive to the spin of the black hole.}
 \label{Fig 4}
\end{figure}

\begin{figure}
\centering{\includegraphics{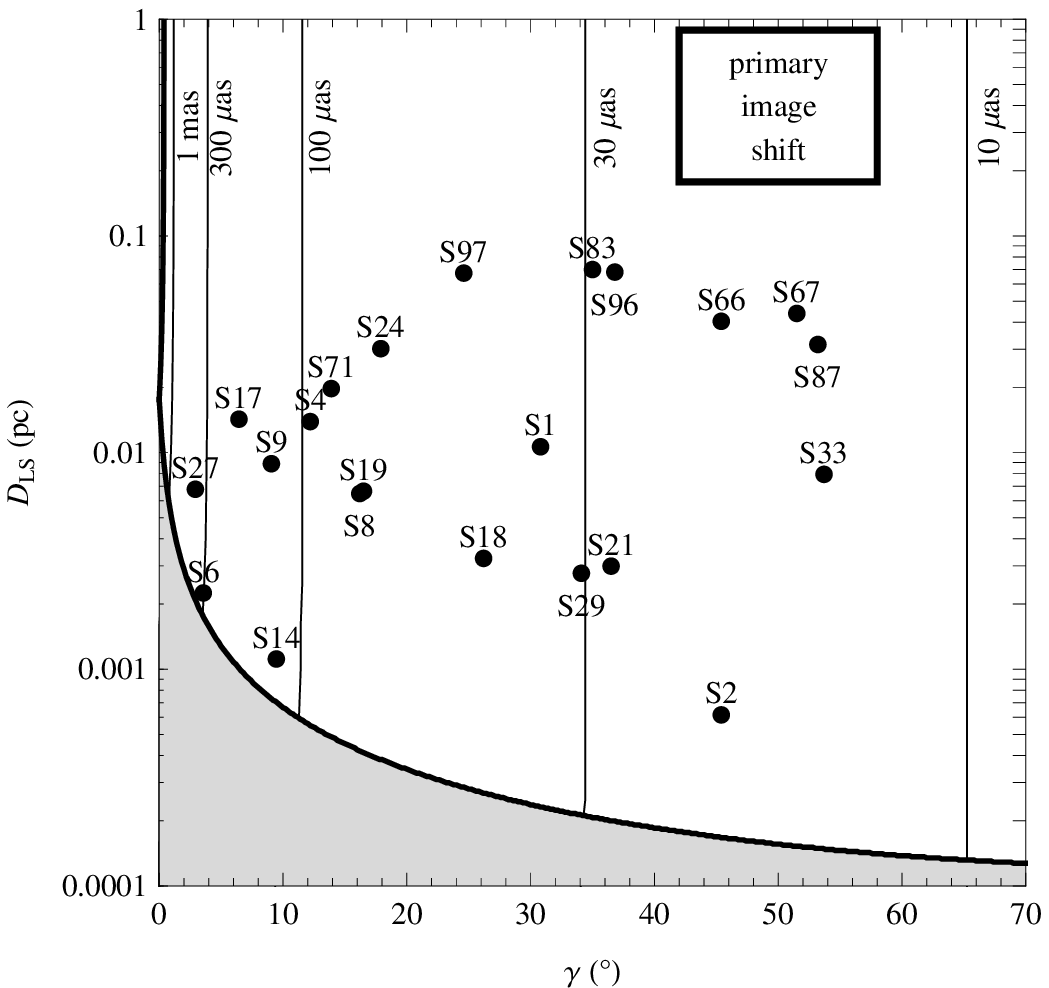}}
 \caption{Details of region II, where both the secondary image of a generic S-star
 is superimposed onto Sgr A*, while the primary is not. The astrometric
 displacement of the primary image due to the gravitational lensing by the central
 black hole is observable by GRAVITY for many S-stars at the time of their periapse.
 The vertical lines indicate how much the primary image of a generic source star
 is shifted by gravitational lensing in that position of the parameter space.}
 \label{Fig 5}
\end{figure}

\begin{figure}
\centering{\includegraphics{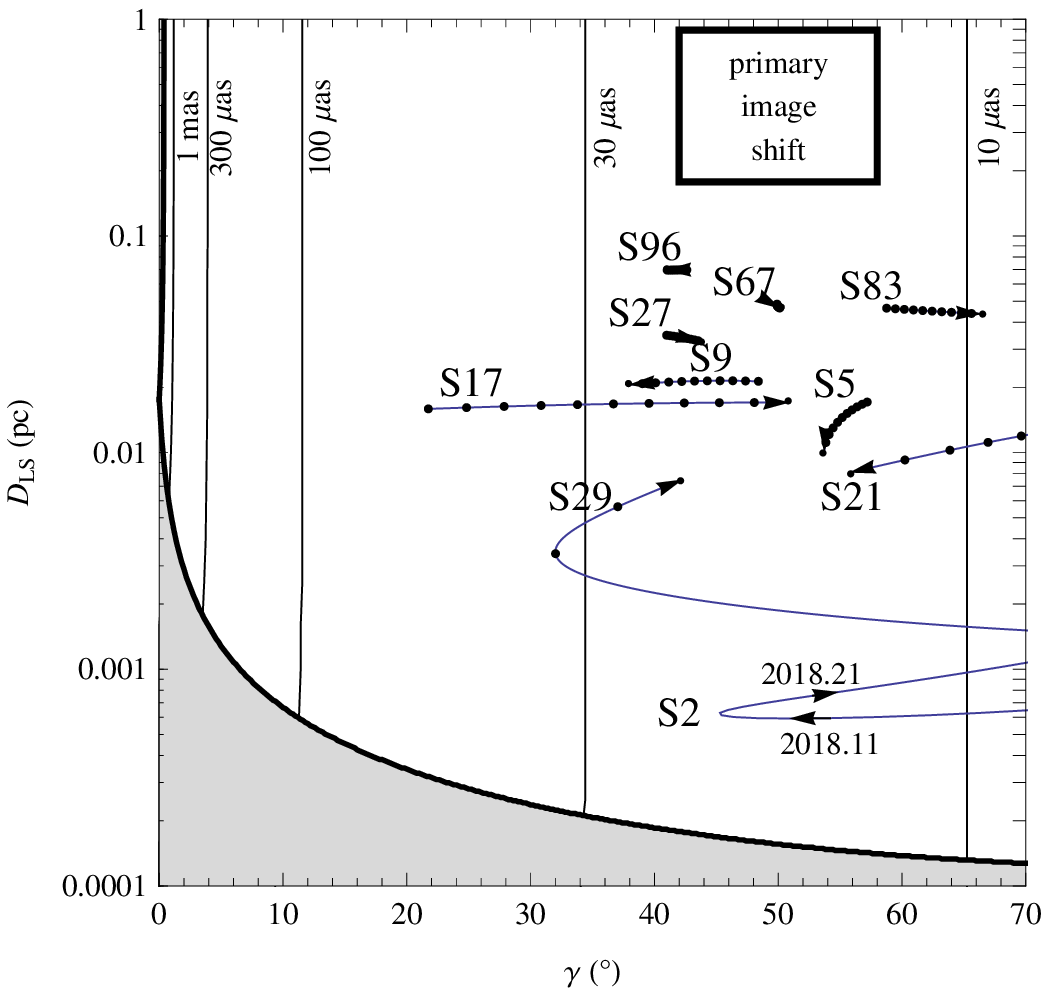}}
 \caption{Details of region II, similar to Figure \ref{Fig 4}. Here the position of several S-stars are reported for the period 2014-2024,
 when GRAVITY is fully operative. The arrows indicate if the stars are approaching or leaving Sgr A*.}
 \label{Fig 6}
\end{figure}

\begin{figure}
\centering{\includegraphics{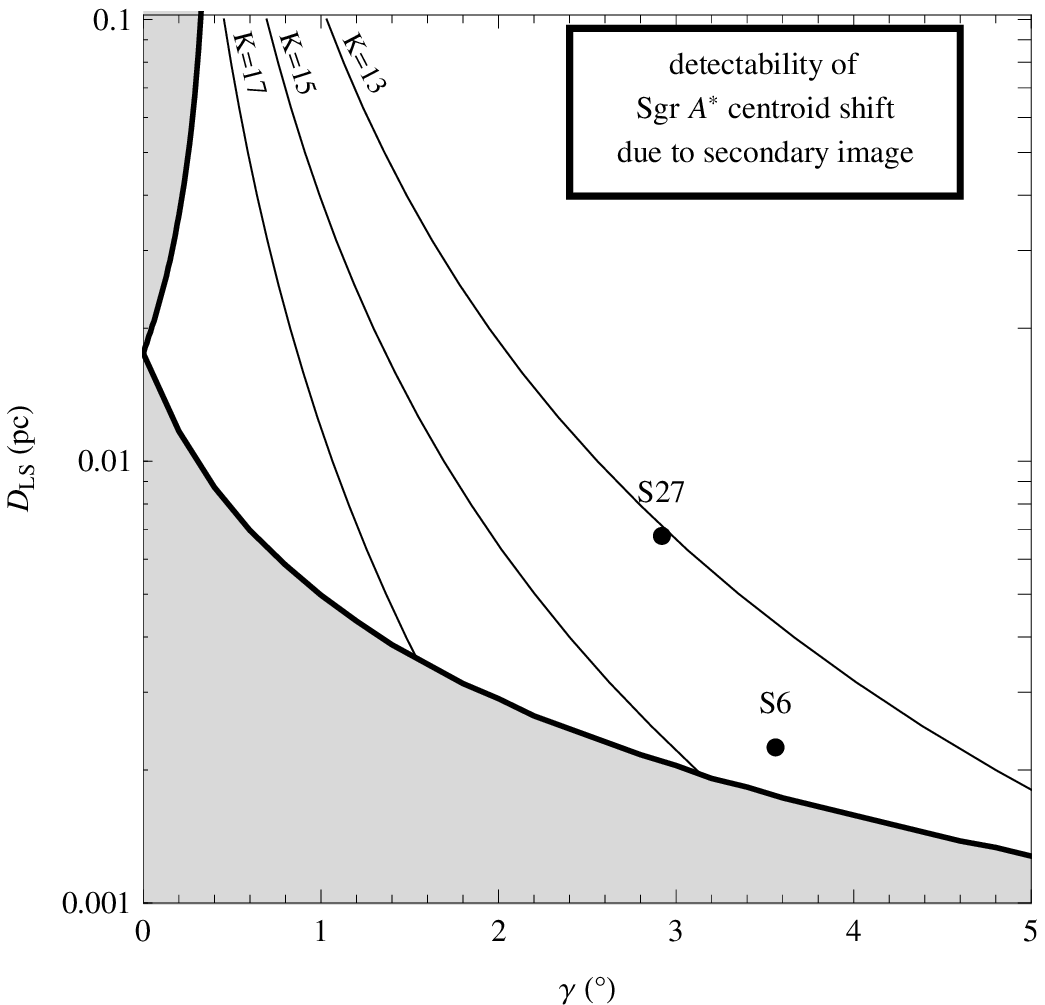}}
 \caption{Details of region II, where the three curves, depending on the magnitude $K$ of the primary source, bound the region in which the centroid shift of Sgr A* due to the secondary image is observable by GRAVITY ($\delta=25\,\mu$as).}
 \label{Fig 7}
\end{figure}

\begin{figure}
\centering{\includegraphics{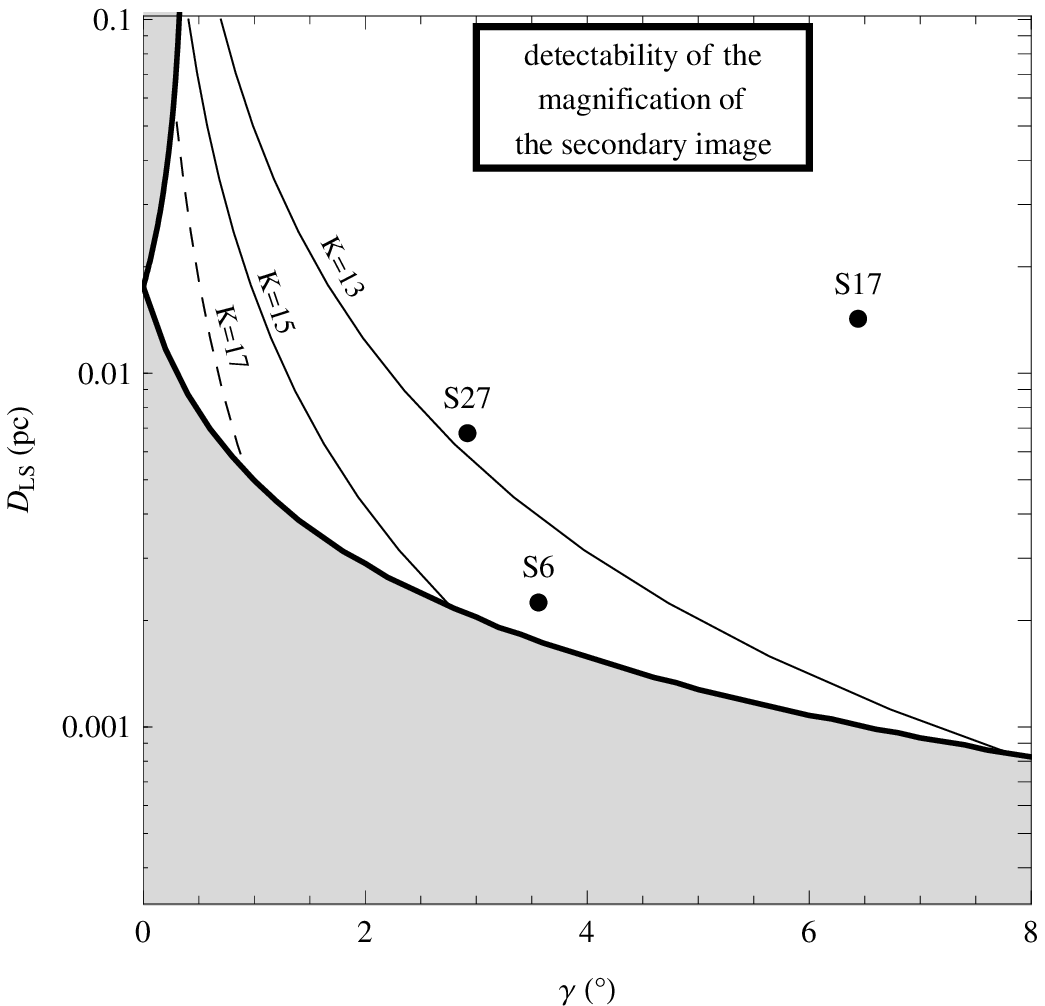}}
 \caption{Details of region II, where the three curves represent the photometric detection limit for the magnification of the secondary image (here blended with Sgr A*), for a threshold of $\Delta K=0.1$. The dashed curve (valid for $K=17$) also represents the detection limit of photometric effects on the primary image.}
 \label{Fig 8}
\end{figure}

\begin{figure}
\centering{\includegraphics{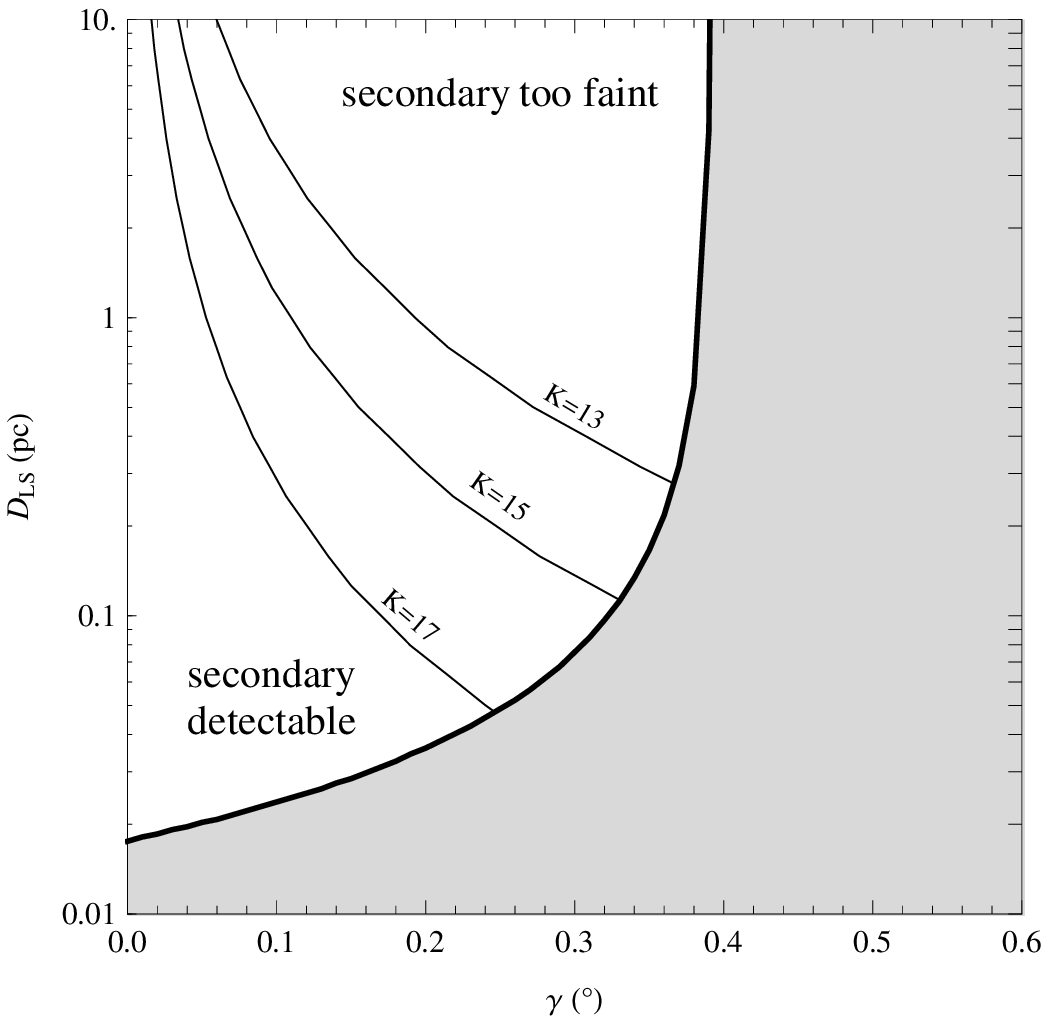}}
 \caption{Details of region III, where the three curves (for three values of the source magnitude) bound the region in which a source must fall in order to have a secondary image brighter than $K=19$.
}
 \label{Fig 9}
\end{figure}

\begin{itemize}

\item[\textbf{Region I}] %
For stellar sources in the immediate neighborhood of Sgr A*, both the primary and the secondary images are confused with Sgr A*.

Gravitational lensing is responsible of a shift in the centroid with respect to the unlensed configuration. This shift is measurable by GRAVITY up to an alignment angle of $\gamma\sim 30^\circ$ (cfr. Figure \ref{Fig 3}, see Figure \ref{Fig 1} for the geometrical definition of $\gamma$).

Photometric detection of a magnification effect is possible in the inner region $\gamma<2 \div 5^\circ$ (Figure \ref{Fig 4}).

These effects have timescales much longer than the daily variability of Sgr A*, which could be subtracted by a dense sampling.

\item[\textbf{Region II}] %
If the S-star is far enough from the black hole, the primary image is resolved by GRAVITY. All known S-stars fall in this region.

In this case, the astrometric shift of the primary is appreciable by GRAVITY for many known S-stars (Figs. \ref{Fig 5}-\ref{Fig 6}) in a cone with aperture $\gamma \sim 40^\circ$. This shift must be taken into account in the computation of orbital parameters and may help constraining the mass of the black hole and of its environment. The star S17, in particular, will have the highest astrometric shift for several years.

The magnification effect on the primary is more difficult to see and requires $\gamma<1^\circ$ (Figure \ref{Fig 8}).

The secondary image is blended with Sgr A*, but is able to yield a measurable centroid shift in the zone between $1^\circ<\gamma<8^\circ$ depending on the source distance and magnitude (Figure \ref{Fig 7}).

The magnification effect on the secondary image might be easier to detect (Figure \ref{Fig 8}), given the relatively low luminosity of Sgr A* in its quiescent state, but it is affected by the problem of the variability of the comntaminant emission, as mentioned before.

\item[\textbf{Region III}] %
In a cone of aperture $\gamma=0.39^\circ$, the stars will generate two images well separated from Sgr A*. Of course this is the best situation for investigations using  gravitational lensing to determine masses and distances in the Sgr A* environment. By imposing that the secondary image is not too faint, we further restrict the zone suitable for such studies (Figure \ref{Fig 9}).

\end{itemize}

For sources exceptionally aligned on the optical axis and close enough to the black hole, the magnification is sensitive to the spin of the black hole (Figure \ref{Fig 4}). Apart from this case, spin and Post-Newtonian corrections to the weak field lensing are generally below the sensitivity of GRAVITY in all examined regions. However, the corrections to the astrometric shifts in Region III and the shift of the secondary image in Region II are not far from GRAVITY sensitivity and may be extracted through a careful statistical analysis on a large number of observations.

In conclusion, we can say that GRAVITY will be able to provide the first evidences of gravitational lensing effects by Sgr A*, thanks to its striking astrometric abilities. The astrometric shift due to gravitational lensing must be taken into account in the reconstruction of the orbits of S-stars and can provide independent precious information on the physics of the massive black hole at the center of our Galaxy. New exciting windows on effects beyond the weak field gravity might also be opened if more S-stars are discovered.

%
%

%

\acknowledgments %
We wish to thank Stefan Gillessen for useful
discussion about the photometric accuracy of NIR measurements.

\end{document}